\documentclass[notitlepage,preprint]{article}
\usepackage[utf8]{inputenc} % Acepta caracteres en castellano
\usepackage[bitstream-charter]{mathdesign} %nice
\usepackage[colorlinks=true,urlcolor=blue,linkcolor=blue, citecolor=blue]{hyperref}
\usepackage{graphicx}
\usepackage[a4paper,left=3.5cm, right=3.5cm, top=3cm, bottom=3cm]{geometry}
\usepackage{authblk}

\usepackage{amsthm}
\usepackage{amsmath}

\title{Geant4 based simulation of the Water Cherenkov Detectors of the LAGO Project}

\author[1]{Rolando Calder\'on}
\author[2,1]{Hernan Asorey\thanks{Corresponding author: \href{mailto://hasorey@uis.edu.co}{hasorey@uis.edu.co}}}
\author[1]{Luis A. N\'u\~nez}

\affil[1]{Escuela de F\'isica, Universidad Industrial de Santander,
Bucaramanga, Colombia}
\affil[2]{Laboratorio Detecci\'on de Part\'iculas y Radiaci\'on, Centro At\'omico Bariloche \& Instituto Balseiro, San Carlos and Bariloche, Argentina}

\date{\today}
\begin{document}
\maketitle

\begin{abstract}
%% Text of abstract
To characterize the signals registered by the different types of water
Cherenkov detectors (WCD) used by the LAGO (Latin American Giant Observatory)
Project, it is highly necessary to develop detailed simulations of the detector
response to the flux of secondary particles at the detector level.  These
particles are originated during the interaction of cosmic rays with the
atmosphere. In this context, the LAGO project aims to study the high energy
component of gamma rays bursts (GRBs) and space weather phenomena by looking
for the solar modulation of galactic cosmic rays (GCRs). Focus in this, a
complete and complex chain of simulations is being developed that account for
geomagnetic effects, atmospheric reaction and detector response at each LAGO
site. In this work we shown the first steps of a GEANT4 based simulation for
the LAGO WCD, with emphasis on the induced effects of the detector internal
diffusive coating.
\end{abstract}

%%%%%%%%%%%%%%%%%%%%%%%%%%%%%%%%%%%%%%%%%%%%%%%
\section{The Water Cherenkov Detectors of the LAGO Project}
%%%%%%%%%%%%%%%%%%%%%%%%%%%%%%%%%%%%%%%%%%%%%%%

During the interaction with the atmosphere, cosmic rays produce large cascades
of secondary particles called extensive air showers (EAS). The spatial and
temporal distributions of these secondary particles at ground had been used for
decades in several observatories to study the underlying mechanisms in the
development of these cascades. To get a deeper understanding of this phenomena,
the characterization of the detector response is crucial. Several techniques
had been used at different experiments, such as complementary measurements with
different type of detectors, first principles calculations and detailed
simulations.

\begin{figure}[t] \centering
  \noindent{\centering\includegraphics[width=0.70\columnwidth]{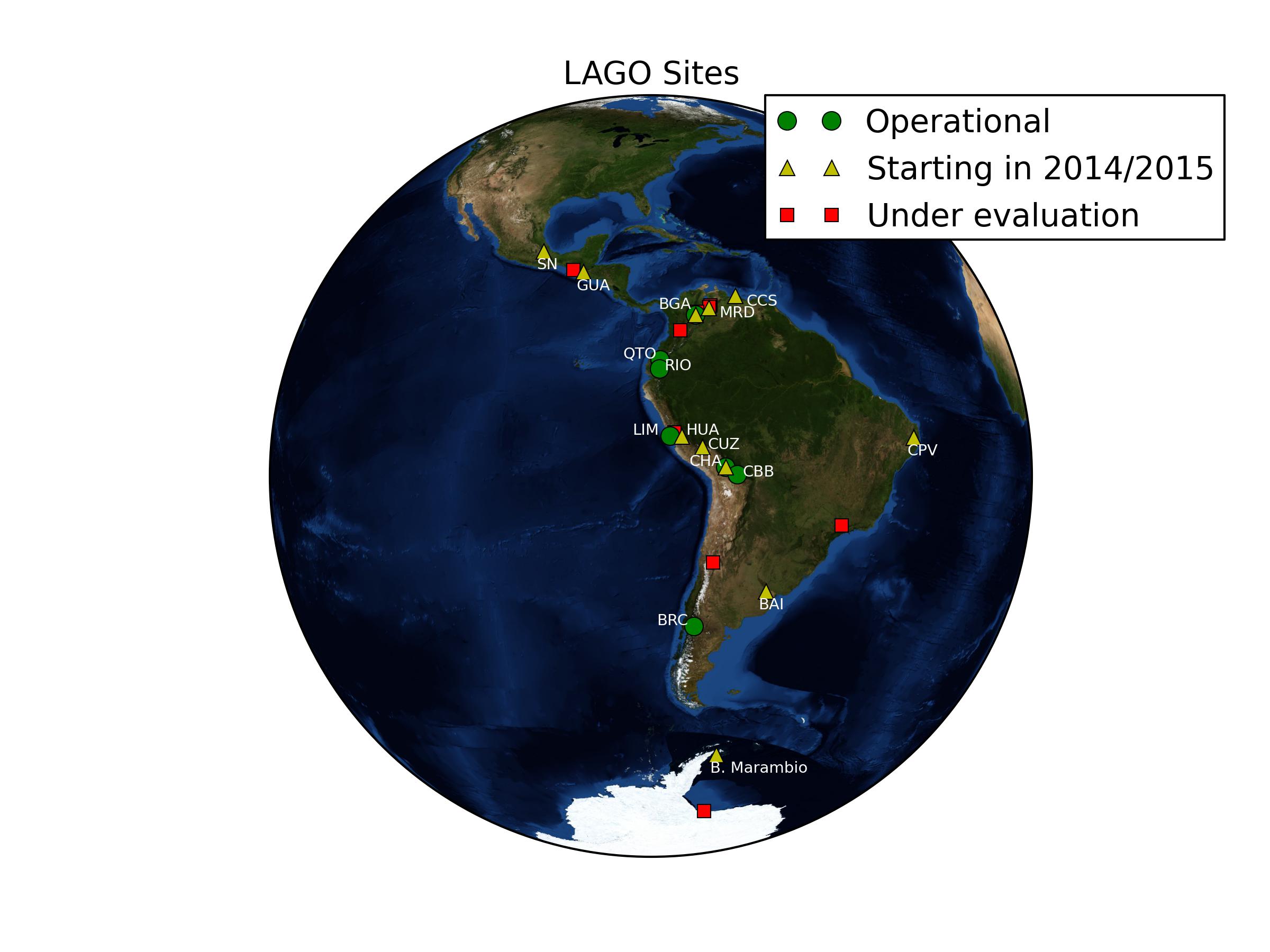}}
  \vspace{-0.2cm} \caption{Current status of the LAGO detectors network in
  Latin America (green circles: operative sites; yellow triangles:
  starting/started in 2014-2015; and red squares: sites under
  evaluation).}\label{lago_sites}
\end{figure}

The Latin American Giant Observatory (LAGO, formerly know as Large Aperture GRB
Observatory) is an extended cosmic ray observatory composed by a network of
water Cherenkov detectors (WCDs) spanning over different sites located at
significantly different latitudes (from the south of Mexico up to the antarctic
region) and different altitudes (from sea level up to more than $5000$\,m
a.s.l.). This network covers a huge range of geomagnetic rigidity cut-offs and
atmospheric absorption/reaction levels. In the figure \ref{lago_sites} we show
the current status of the detection network, designed to measure with extreme
detail the temporal evolution of the radiation flux at ground level. It is
mainly oriented to perform basic research on three branches: the extreme
universe, space weather phenomena and atmospheric radiation at ground
level\,\cite{Allard2008,Asorey2013b}. Several scientific and academic programs
are conducted within the LAGO framework in Latin America\,\cite{Asorey2014a}.

LAGO is being built and is operated by the LAGO Collaboration, a
non-centralized collaborative union of more than 30 institutions from nine
Latin American countries. Using observations from our WCDs, it is possible to
study the galactic modulation of galactic cosmic rays from combining different
ground sites, in particular it is possible to study the long-term modulation as
well as transient events. 

\begin{figure}[t] \centering
	\noindent{\centering\includegraphics[width=0.7\columnwidth]{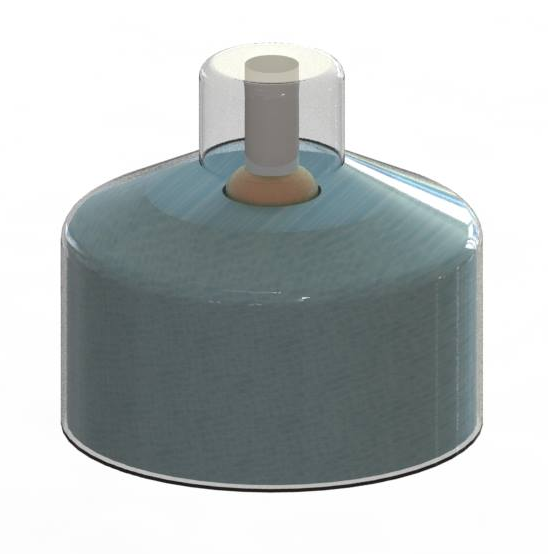}}
	\vspace{-0.2cm} 
	\caption{Typical water Cherenkov detector of the LAGO project detection
		network. It is composed by a commercial water tank, a single central
		photomultiplier tube and an internal coating made by a highly diffusive
		and reflective material.} 
	\label{lago_wcd} 
\end{figure}

The LAGO water Cherenkov detectors are built starting from commercial water
tanks with $\sim 1$\,m$^3$ to $40$\,m$^3$ of purified water. The passage of
ultra-relativistic charged particles through the water volume produce Cherenkov
light that is collected by a central photomultiplier tube (PMT), typically an
8-inch Hamamatsu R5912 PMT. The sensitivity to secondary gammas in the cascade
is enabled by the production of Cherenkov capable electron/positron pairs
within the water volume. The detector has an internal coating made by a highly
diffusive and reflective fabric of commercial
Tyvek\textregistered
%\footnote{http://bit.ly/1whxoph}
. The diffusion of
Cherenkov photons in the fabric surface reduce the signal dependence with the
secondary particle trajectory within the detector. A FPGA based, own designed
fast analog-to-digital conversion electronics allows the operation of up to
four independent detectors in a same site. This electronics is controlled by a
Raspberry Pi or similar device. Additionally, the station has a temperature and
atmospheric pressure sensor and a GPS board for time synchronization between
different LAGO sites. The power consumption of the station is less than $11$\,W
and is powered by solar panels and batteries. The LAGO WCD are characterized by
its low cost and reliability, and a schema of this detectors can be seen in
figure \ref{lago_wcd}.

%%%%%%%%%%%%%%%%%%%%%%%%%%%%%%%%%%%%%%%%%%%%%%
\section{Detector simulation}
%%%%%%%%%%%%%%%%%%%%%%%%%%%%%%%%%%%%%%%%%%%%%%

When a particle travels in a medium at a faster speed than the
wavelength-dependent phase velocity of the EM field in that medium, it emits
Cherenkov radiation at those wavelengths, a phenomenon which is totally
independent of Bremsstrahlung. It can be seen that the number of Cherenkov
photons by unit of interaction depth $X$ that are radiated by the medium with
refractive index $n(\lambda)$ in the wavelength interval $\lambda_1 \leq \lambda
\leq \lambda_2$ is given by:
\begin{equation}
  \frac {\mathrm{d}N }{\mathrm{d}X}= 2 \pi \alpha \left ( 1- \frac {1}{ { \beta
  }^{ 2 }{ n }^{ 2 }(\lambda ) }  \right) \left( \frac { 1 }{ { \lambda  }_{ 2
  } } -\frac { 1 }{ { \lambda  }_{ 1 } }  \right),
\label{equation2}
\end{equation}
where $\alpha =(e^2/{\hbar c})$ is the fine structure constant. In the typical
wavelength range of interest for commercial PMTs, an ultra-relativistic charged
particle produce $\sim 30$ Cherenkov photons per millimeter in pure water.

The Geant4 framework\,\cite{Pia2003} provides internal routines to generate
Cherenkov photons based on the formula above. For each charged particle
propagating through a medium it verifies the Cherenkov condition and produce
photons with the corresponding wavelength distribution. 

The detector simulated in this work correspond to one of the operating
detectors at the Universidad Industrial de Santander (UIS), located in the city
of Bucaramanga, Colombia, at $956$\,m a.s.l. This detector, Guane-3, was built
with a commercial water tank of radius $r=0.515$\,m and total height
$h=0.59$\,m. The PMT was simulated as an hemispheric volume of radius
$r_P=0.101$\,m immersed in the water volume at the roof centre. Each Cherenkov
photon impinging on the PMT surface was counted with a constant probability of
$0.25$ (independent of the photon wavelength), which correspond to the
approximated quantum efficiency (QE) of the Hamamatsu R5912 PMT in the range
$330-570$\,nm. Cherenkov photons with wavelength out of this range were simply
discarded. For the medium we consider pure water with refractive index
depending on the photon wavelength.

The inner surface coating was simulated following the guidelines given by
Janecek and Moses\,\cite{Janecek2010}: at the code level, the routine
representing the internal surface has a pointer to a table describing a
particular surface. If the surface is, e.g., painted, wrapped, or has a
cladding, the table include a multiple layer with different refractive index.
The Tyvek diffusive properties are simulated by using look-up tables that
describe the inner surface properties and were constructed from experimental
measures from Filevich et al.\,\cite{Filevich1999}

\begin{figure}[t]
	\centering 
	\noindent{\centering\includegraphics[width=0.70\columnwidth]{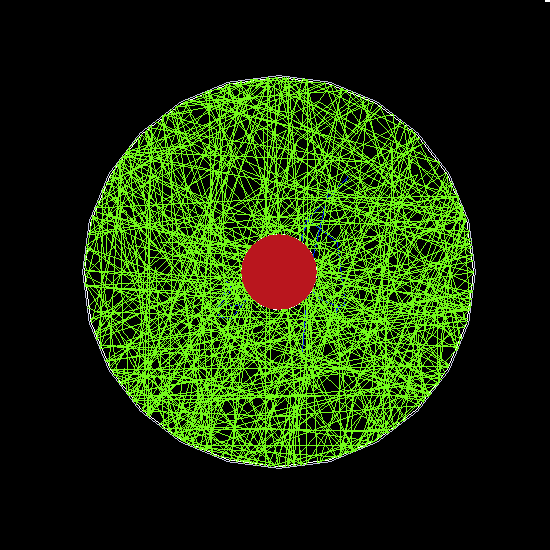}}
	\vspace{-0.2cm}
	\caption{Cherenkov photons (green lines) produced during the propagation of
	a vertical muon with energy $E=100$\,MeV across the simulated detector
	volume. The simulated PMT is visible as a red mesh in the central part of
	the detector roof.}
	\label{muon1}
\end{figure}

Finally vertical muons with different energies were injected trough the water
volume. As an example, in figure \ref{muon1} we show the Cherenkov photons
produced by a muon with energy $E_\mu=100$\,MeV. In this case, the muon has
enough energy to completely cross along the detector.

\begin{figure}[t]
	\centering 
	\noindent{\centering\includegraphics[width=0.8\columnwidth]{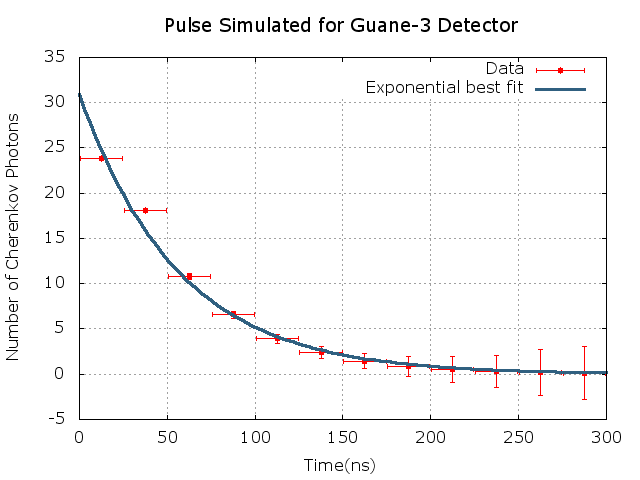}} 
	\vspace{-0.2cm}
	\caption{Simulated pulse shape produced by the passage trough the UIS
		detector of an atmospheric muon with energy $E=5$\,GeV. The
		characteristic time obtained by fitting an exponential function (solid 
		line) is $\tau=(55.1\pm0.7)$\,ns, which is consistent with the observed
		pulse at the detector.}
	\label{muon2}
\end{figure}

It can be deduced from equation (\ref{equation2}) that the number of Cherenkov
photons produced by the passage of a particle trough the detector only depends
of the medium (water) and of the particle speed ($\beta c$). Moreover, the
number of photons $\mathrm{d}N/\mathrm{d}X$ in a fixed range of wavelengths is
nearly constant for large $\beta$s. Bearing in mind the properties of the
diffusive material, and for a fixed detector geometry, it is clear then that
the pulse shape will depend mainly in the range of each type of particle in the
water volume. So we use the detector pulse shape, i.e. the time distribution of
Cherenkov photons impinging on the PMT, as a clear indicator of the detector
response for each type of particle. In the figure \ref{muon2} we show the
averaged pulse shape produced by $200$ individual vertical muon with energy
$E=5$\,GeV for the WCD typical internal coating. By fitting an exponential
function of the form $f(t)=N_0 \exp \left ( -t / \tau \right )$, we obtained
that the characteristic time of the pulse produced by a vertical atmospheric
muon in Guane-3 is $\tau=(55.1\pm0.7)$\,ns, which is consistent with the
observed pulse in the detector in the signal range corresponding to vertical
muons.  

%%%%%%%%%%%%%%%%%%%%%%%%%%%%%%%%%%%%%%%%%%%%%%
\section{Conclusions and Acknowledgements}
%%%%%%%%%%%%%%%%%%%%%%%%%%%%%%%%%%%%%%%%%%%%%%

In this work we show the first results of a completely based GEANT4 simulation
of one of the LAGO WCD installed at the Universidad Industrial de Santander, in
Bucaramanga, Colombia. By using the pulse shape as an indicator of the detector
response, we obtained the averaged pulse shape produced by a vertical
atmospheric muon traversing the detector volume, with a characteristic time
$\tau = (55.1\pm0.7)$, consistent with the real muon pulse of this detector.   

The authors of this work thank the support of COLCIENCIAS ``Semillero de
Investigaci\'on'' grant 617/2014 and
CDCHT-ULA project C-1598-08-05-A.

%%%%%%%%%%%%%%%%%%%%%%%%%%%%%%%%%%%%%%%%%%%%%%%
%% References with BibTeX database:
\bibliographystyle{elsarticle-num}
\bibliography{biblio}

\begin{thebibliography}{1}
\expandafter\ifx\csname url\endcsname\relax
  \def\url#1{\texttt{#1}}\fi
\expandafter\ifx\csname urlprefix\endcsname\relax\def\urlprefix{URL }\fi
\expandafter\ifx\csname href\endcsname\relax
  \def\href#1#2{#2} \def\path#1{#1}\fi

\bibitem{Allard2008}
D.~Allard, I.~Allekotte, C.~Alvarez, H.~Asorey, et~al., {Use of water-Cherenkov
  detectors to detect Gamma Ray Bursts at the Large Aperture GRB Observatory
  (LAGO)}, Nuclear Instruments and Methods in Physics Research Section A:
  Accelerators, Spectrometers, Detectors and Associated Equipment 595~(1)
  (2008) 70--72.
\newblock \href {http://dx.doi.org/10.1016/j.nima.2008.07.041}
  {\path{doi:10.1016/j.nima.2008.07.041}}.

\bibitem{Asorey2013b}
{The LAGO Collaboration [H. Asorey]}, {The LAGO Solar Project}, in: Proceedings
  of the 33th International Cosmic Ray Conference ICRC 2013, Vol. in press,
  R\'{\i}o de Janeiro, Brazil, 2013, pp. 1--4.

\bibitem{Asorey2014a}
H.~Asorey, {The Latin American Giant Observatory Project}, in: Proceedings of
  the X SILAFAE, Medell\'in, Colombia, 2014.

\bibitem{Pia2003}
M.~Pia, {The Geant4 Toolkit: simulation capabilities and application results},
  Nuclear Physics B - Proceedings Supplements 125 (2003) 60--68.
\newblock \href {http://dx.doi.org/10.1016/S0920-5632(03)90967-4}
  {\path{doi:10.1016/S0920-5632(03)90967-4}}.

\bibitem{Janecek2010}
M.~Janecek, W.~W. Moses, Simulating scintillator light collection using
  measured optical reflectance, Nuclear Science, IEEE Transactions on 57~(3)
  (2010) 964--970.

\bibitem{Filevich1999}
A.~Filevich, P.~Bauleo, H.~Bianchi, J.~R. Martino, G.~Torlasco,
  Spectral-directional reflectivity of tyvek immersed in water, Nuclear
  Instruments and Methods in Physics Research Section A: Accelerators,
  Spectrometers, Detectors and Associated Equipment 423~(1) (1999) 108--118.

\end{thebibliography}
\end{document}